\documentclass{aa}  

\usepackage{color}
\usepackage{graphicx}
\usepackage{txfonts}
\begin{document} 

   \title{Spectral evolution of the supergiant HMXB IGR~J16320--4751 along its orbit using XMM-{\em Newton}}

\titlerunning{Spectral evolution of sgHMXB IGR~J16320--4751 using XMM-{\em Newton}}

   \author{Federico Garc\'{\i}a\inst{1,2,3}
          \and
          Federico A. Fogantini\inst{2,3}
          \and
          Sylvain Chaty\inst{1}
          \and
          Jorge A. Combi\inst{2,3}
          }

   \institute{
   Laboratoire AIM (UMR 7158 CEA/DRF-CNRS-Universit\'e
Paris Diderot), Irfu/D\'epartement d’Astrophysique, Centre de Saclay, FR-91191 Gif-sur-Yvette Cedex, France
\and
   Instituto Argentino de Radioastronom\'{\i}a (CCT-La Plata, CONICET; CICPBA), C.C. No. 5, 1894 Villa Elisa, Argentina
                \and
Facultad de Ciencias Astron\'omicas y Geof\'{\i}sicas, Universidad Nacional de La Plata, Paseo del Bosque s/n, 1900 La Plata, Argentina\\
              \email{federico.garcia@cea.fr, fgarcia@iar.unlp.edu.ar, fgarcia@fcaglp.unlp.edu.ar}
         }

   \date{Received ***; accepted ***}

  \abstract
{The {\em INTEGRAL} satellite has revealed a previously hidden population of absorbed high-mass X-ray binaries (HMXBs) hosting supergiant stars. Among them, IGR~J16320--4751 is a classical system intrinsically obscured by its environment, with a column density of $\sim$10$^{23}$~cm$^{-2}$, more than an order of magnitude higher than the interstellar absorption along the line of sight. It is composed of a neutron star rotating with a spin period of $\sim$1300~s, accreting matter from the stellar wind of an O8I supergiant star, with an orbital period of $\sim$9~days.}
{We  investigated the geometrical and physical parameters of both components of the binary system IGR~J16320--4751. Since in  systems of this type the compact object is usually embedded in the dense and powerful wind of an OB supergiant companion, our main goal here was to study the dependence of the X-ray emission and column density along the full orbit of the neutron star around the supergiant star.}
{We analyzed all existing archival XMM-{\em Newton} and {\em Swift}/BAT observations collected between 2003 and 2008, performing a detailed temporal and spectral analysis of the X-ray emission of the source. We then fitted the parameters derived in our study, using a simple model of a neutron star orbiting a supergiant star.}
{The XMM-{\em Newton} light curves of IGR~J16320--4751 display high-variability and flaring activity in X-rays on  several timescales, with a clear spin period modulation of $\sim$1300~s. In one observation we detected two short and bright flares where the flux increased by a factor of $\sim$10 for $\sim$300~s, with similar behavior in the soft and hard X-ray bands. By inspecting the 4500-day light curves of the full {\em Swift}/BAT data, we derived a refined period of 8.99$\pm$0.01~days, consistent with previous results. The XMM-{\em Newton }spectra are characterized by a highly absorbed continuum and an Fe absorption edge at $\sim$7~keV. We fitted the continuum with a thermally comptonized {\sc comptt} model, and the emission lines with three narrow Gaussian functions using two {\sc tbabs} absorption components, to take into account both the interstellar medium and the intrinsic absorption of the system. For the whole set of observations we 
derived the column density at different orbital phases, showing that there is a clear modulation of the column density with the orbital phase. In addition, we also show that the flux of the Fe K$\alpha$ line is 
correlated with the N$_{\rm H}$ column, suggesting a clear link between absorbing and fluorescent matter that, together with the orbital modulation, points towards the stellar wind being the main contributor to 
both continuum absorption and Fe~K$\alpha$ line emission.}
{Assuming a simple model for the supergiant stellar wind we were able to explain the orbital modulation of the absorption column density, Fe~K$\alpha$ emission and the high-energy {\em Swift}/BAT flux, allowing us to constrain the geometrical parameters of the binary system. Similar studies applied to the analysis of the spectral evolution of other sources will be useful to better constrain the physical and geometrical properties of the sgHMXB class.}


   \maketitle

\section{Introduction}

Since its launch in 2002, the IBIS/ISGRI detector \citep{lebrun2003,ubertini2003} on board the International Gamma-Ray Astrophysics Laboratory ({\em INTEGRAL}) has discovered several sources belonging to a class of highly absorbed low-luminosity X-ray sources, which makes them difficult to detect in the soft X-ray band ($<$3~keV). Most of them exhibit high levels of obscuration \citep[e.g.,][]{zuritaheras2006,rodriguez2006} or extreme flaring behavior characterized by hard X-ray flux variations of several orders of magnitude on timescales of a few hours \citep{negueruela2006,sguera2006}. The great majority of this new kind of hard X-ray sources are of Galactic origin since {\em INTEGRAL} has spent a considerable fraction of its observational time pointing towards the Galactic plane. Observations performed with missions such as XMM-{\em Newton} or {\em Chandra}, with high astrometric accuracy at the level of the arc-second, allowed the determination of their position leading to the possible optical identification of their companion counterparts \citep{chaty2008,coleiro2013}. For instance, more than a dozen of these X-ray sources were detected in the direction of the Norma Galactic spiral arm and the vast majority of them are thought to be high-mass X-ray binaries (HMXB) with early-type companion stars. Many of these obscured HMXBs were found to be persistent for several years showing high intrinsic variability on several timescales. In some cases a regular period was detected in the hard X-ray band, generally attributed to the orbital motion \citep{corbet1986}.

The transient hard X-ray source IGR~J16320--4751 belongs to this new class of highly absorbed binary systems. The source was discovered on February 1, 2003, with the {\em INTEGRAL} observatory during ToO observations of 4U~1630-47 \citep{tomsick2003}. It showed a significant variability in the 15$-$40~keV energy range, being also detected in some occasions above 60~keV \citep{tomsick2003,foschini2004}. Its position RA$_{\rm J2000}$ = 16h32m DEC$_{\rm J2000}$ = --47d51m was coincident with the X-ray source AX~J1631.9$-$4752, which was previously observed with the ASCA telescope \citep{sugizaki2001}. The ASCA spectrum was modeled by a power law with a hard photon index $\Gamma \sim 0.2\pm0.2$ \citep{sugizaki2001}, which suggested that the source could belong to the HMXB class. 

Follow-up observations of the source with XMM-{\em Newton} on March 4, 2003, confirmed the complex temporal behavior of the source, showing several flaring events without significant variations in its hardness \citep{rodriguez2003}. These observations  improved the source position to a radius of 3~arcsec of accuracy at RA=16h32m01.9s DEC=-47$^\circ$52'27'' \citep{rodriguez2003,rodriguez2006}.
Using the same XMM-{\em Newton} observations, \cite{lutovinov2005} were able to unambiguously identify a pulsation period of $P=1309\pm40$~sec, confirming the \cite{rodriguez2003} claim about the nature of the system as an HMXB with an accreting neutron star (NS). 
Later on, using {\em INTEGRAL} observations, \cite{rodriguez2006} were able to confirm the presence of pulsations above 20~keV, clearly present in the Comptonized emission realized in the close vicinity of the NS, with a pulse fraction independent of the energy. Combining the XMM-{\em Newton} and {\em INTEGRAL} spectra, \cite{rodriguez2006} fitted an absorbed power law  with $\Gamma \sim 1.6$ and a high absorption column of $N_{\rm H} \sim 2.1 \times 10^{23}$~cm$^{-2}$, incorporating a narrow iron line at $\sim$6.4~keV.
An orbital period of 8.96$\pm$0.01 days was found from a {\em Swift}/BAT light curve extending from  December 21, 2004, to  September 17, 2005 \citep{corbet2005}, and of 8.99$\pm$0.05 days with {\em INTEGRAL} \citep{walter2006}.  The location of this system on the pulse/orbital period diagram \citep{corbet1986} is typical of a NS accreting from the wind of an early supergiant companion.

\cite{chaty2008} identified the most likely infrared counterpart \citep[2MASS J16320215--4752289, invisible in the optical,][]{rodriguez2006}, in agreement with \cite{negueruela2007}, and rejecting other possible candidates based on a photometric analysis. In their blue near-infrared (NIR) spectra, \cite{chaty2008} detected only a few lines due to the high absorption, while in their red NIR spectrum they found several absorption and emission lines, such as the Pa(7-3) emission line, the Brackett series with P Cygni profiles between 1.5 and 2.17~$\mu$m, and He~I at 2.166~$\mu$m, leading to the classification of the companion star as a luminous supergiant OB star, which helped to unambigously identify the stellar counterpart. Their result was also in agreement with the SED fit computed by \cite{rahoui2008} which, including mid-IR observations, derived an optical absorption $A_{\rm v} = 35.4$~mag and an O8I spectral type for the companion with $T \approx 33 000$~K and $R \approx 20$~R$_\odot$, suggesting a distance of 3.5~kpc to the source. Finally, in a more recent study, \cite{coleiro2013} performed NIR spectroscopy showing a faint broad He~I emission and classified the stellar companion as an BN0.5~Ia supergiant star.

In this paper we report a detailed temporal and spectral analysis of nine XMM-{\em Newton} public observations of IGR~J16320--4751. In Sect. 2 we provide details about the XMM-{\em Newton} and {\em Swift}/BAT observations and data reduction methods that were employed for the analysis. We describe the temporal and spectral X-ray analysis and results in Sect. 3. Finally, in Sect. 4 we discuss these results in the context of a simple model developed to account for the spectral orbital variability.

\label{sec_obs}
\section{Observations and data analysis}

\subsection{XMM-{\em Newton} data}

\begin{table*}[]
\centering
\begin{tabular}{c c c c c c c c c}
\hline
OBSID & Start date & End date & Filter/Mode & Exp. Time & GTI & Exc. Radius & Rate & Color \\ 
& [UTC] & [UTC] & & [ks] & [ks] & [arcsec] & [cts~s$^{-1}$] &\\
\hline
0128531101 & 2003-03-04 \,\, 20:58 & 2003-03-05 \,\, 03:12 & Medium/LW & 4.78 & 4.47 & 0 & 0.2 & 0.88  \\
0201700301 & 2004-08-19 \,\, 13:28 & 2004-08-20 \,\, 03:20 & Thin1/FW & 38.0 & 33.9 &10 & 0.4 & 1.09\\
\hline
0556140101 & 2008-08-14 \,\, 22:41 & 2008-08-15 \,\, 01:12 & Thin1/FW & 5.33 & 4.64 & 4& 0.5  & 1.53 \\
0556150201 & 2008-08-16 \,\, 17:38 & 2008-08-16 \,\, 19:52 & Thin1/FW & 1.63 & 1.42 & 0& 1.0  & 1.52 \\
0556140301 & 2008-08-18 \,\, 13:33 & 2008-08-18 \,\, 15:31 & Thin1/FW & 1.23 & 1.07 & 4& 0.55 & 1.24 \\
0556140401 & 2008-08-20 \,\, 07:34 & 2008-08-20 \,\, 10:41 & Thin1/FW & 11.2 & 9.78 & 4& 0.5  & 1.37 \\
0556140501 & 2008-08-21 \,\, 07:02 & 2008-08-21 \,\, 07:39 & Thin1/FW & 1.34 & 1.16 & 6& 0.45 & 1.33 \\
0556140601 & 2008-08-22 \,\, 03:54 & 2008-08-22 \,\, 07:20 & Thin1/FW & 12.4 & 10.8 &10& 0.45 & 1.39 \\
0556140701 & 2008-08-24 \,\, 18:28 & 2008-08-24 \,\, 20:59 & Thin1/FW & 6.03 & 5.21 & 4& 0.4  & 2.54 \\
0556140801 & 2008-08-26 \,\, 13:33 & 2008-08-26 \,\, 16:13 & Thin1/FW & 9.63 & 8.37 & 8& 0.4  & 1.22 \\
0556141001 & 2008-09-17 \,\, 01:25 & 2008-09-17 \,\, 03:31 & Thin1/FW & 4.33 & 3.77 & 0& 0.45 & 1.38 \\ 
\hline
\end{tabular} 
\caption{XMM-{\em Newton} PN observations used in this work. LW and FW modes correspond to PN Large Window and Full Window, respectively. Exposure time and good time intervals (GTI) are shown in ks. The excision radii used in the pile-up treatment for each observation are indicated in arcsec and the columns  ``Rate'' and ``Color''  correspond to the average count rate and {soft/hard} color ratio in the corresponding annular extraction regions.}
\label{obstable}
\end{table*}

The XMM-{\em Newton} observatory has two X-ray instruments on board: the European Photon Imaging Camera (EPIC) and the Reflecting Grating Spectrometers (RGS). EPIC consists of three detectors, two MOS cameras,  MOS1 and MOS2 \citep{turner2001}, and a PN camera \citep{struder2001}, which operate in the 0.3$-$12~keV energy range. RGS is formed of two high-resolution spectrometers working in the 0.3$-$2.0~keV energy band.

IGR~J16320--4751 was observed twice in March 2003 \citep{rodriguez2003,lutovinov2005} and August 2004 \citep{rodriguez2006}, and nine times between August 14 and September 17,  2008 \citep[for a preliminary analysis, see][]{zuritaheras2009}. The first observation was performed with a medium filter in Large Window (LW) mode, while the rest of the exposures were conducted with a thin filter in Prime Full Window observation mode. Since the PN effective area is several times larger than the MOS CCDs, and the latter were not in many of the observations, we present here the analysis of the EPIC PN data set. As RGS covers only the highly absorbed soft band up to $\sim$2.0~keV, we do not use those spectrometers in our analysis. 

We reduced the XMM-{\em Newton} data by means of the Science Analysis System (SAS) version 16.0.0 and the latest calibrations available on June 2017. We obtained event lists from the PN data set after processing the Observation Data Files (ODF) with the {\sc epproc} task. In order to exclude high-background periods we produced background light curves, excluding a circle of 100~arcsec surrounding the bright IGR~J16320--4751 source, for events with energies above 10~keV. Good time intervals (GTI) were obtained excluding intervals 3$\sigma$ above the mean count rate of each light curve. 

\begin{figure*}
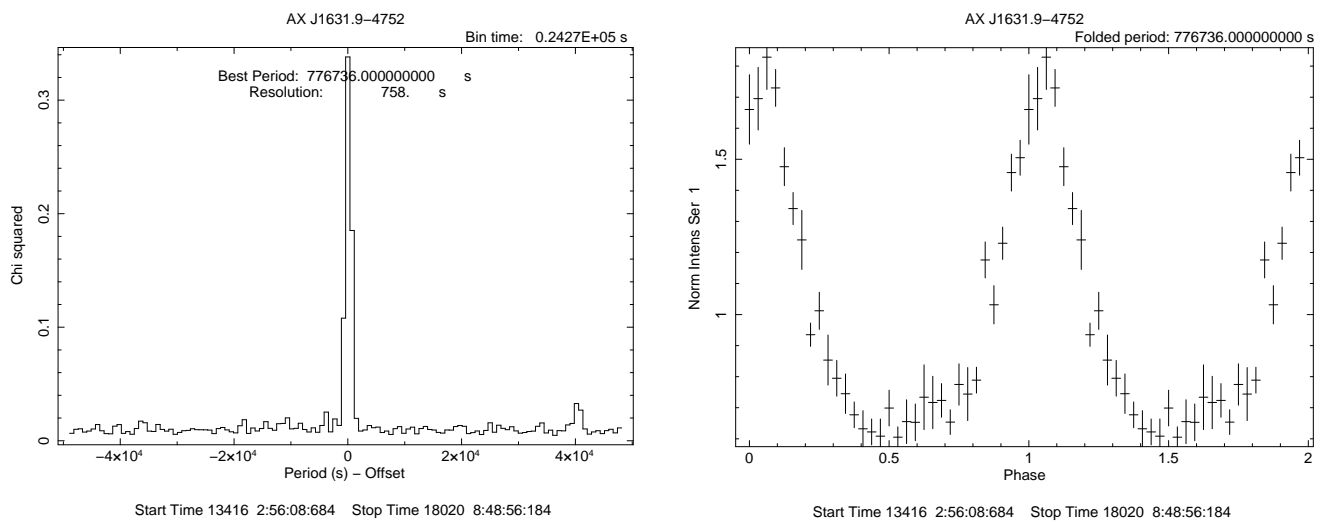

\centering
\includegraphics[angle=-90,width=0.48\hsize]{EFSEARCH_P8.99d.ps}
\includegraphics[angle=-90,width=0.48\hsize]{FOLDED_LIGHTCURVE_BINNED_P8.99d.ps}
   \caption{{\bf Left panel:}  Best period of 8.99~days  found using the {\sc efsearch} of  {\sc heasoft} for the full $\sim$4500-day {\em Swift}/BAT X-ray light curve. {\bf Right panel:} Folded light curve of {\em Swift} BAT data spanning from 2004 Apr 24 to 2017 Sept 26 using the best period.}
      \label{swiftbat}
\end{figure*}

\label{sec_pileup}
\subsubsection{Pile-up treatment}

We investigated in detail the presence of pile-up, since the average count rate in the EPIC PN camera was close to or above the level at which pile-up effects can become significant for the vast majority of the observations in the whole sample. In order to do this, we used the SAS task {\sc epatplot} to create diagnoses of the relative ratios of single  ({\sc PATTERN==0}) and double  ({\sc PATTERN=[1:4]}) events to study their possible deviation from the standard values expected from the PN camera calibrations. 
By selecting circular regions of 30'' radii surrounding IGR~J16320--4751, we found that seven of the nine series and one of the earlier observations were affected by pile-up.

Then, following the standard procedure suggested by the XMM-{\em Newton} calibration team, for each of the piled-up observations we extracted spectra selecting single and double events on several concentric annuli with fixed outer radius of 35'', varying the inner radii by integer factors of the PSF: 80, 120, 160, and 200 in physical units, which correspond to 4'', 6'', 8'', and 10''. We thus extracted one spectrum for each of the defined regions and carefully inspected each {\sc epatplot} produced in order to define the minimum excision radii needed to avoid pile-up in each of the observations. 

For subsequent analysis, events with {\sc flag==0} were selected with {\sc evselect} task, except for the early ObsID 0128531101, for which we used the {\sc xmmeaep} selector. 
In Table~\ref{obstable} we present the whole set of XMM-{\em Newton} observations used throughout this work, including excision radii used, average count rate, and mean color.

In order to check the quality of our pile-up treatment, we produced spectra for single events ({\sc PATTERN==0}), double events ({\sc PATTERN IN [1:4]}), and for a combination of single and double events ({\sc PATTERN<=4}). We fitted them separately and checked that the results were consistent with each other within the uncertainties. After that successful step, for the subsequent analysis we only considered the latter set which had the best S/N.

\subsection{{\em Swift}/BAT data}
\label{ephemer}

The {\em Swift}/Burst Alert Telescope (BAT) is a transient monitor that provides permanent coverage of the hard X-ray sky in the 15--50~keV energy range. The BAT observes $\sim$90\% of the sky each day with a detection sensitivity of 5.3 mCrab in a full-day observation, with a time resolution of 64~s \citep{krimm2013}. The primary interface for the BAT transient monitor is a public website\footnote{https://swift.gsfc.nasa.gov/results/transients/} where more than 900 source light curves are available spanning more than eight years. Between 2005 and 2013, the monitor detected 245 sources: 146 of them being persistent and 99 seen only in outburst.

We used the full {\em Swift}/BAT data available up to September 26, 2017, in the online service\footnote{https://swift.gsfc.nasa.gov/results/transients/AXJ1631.9-4752/} of daily and orbital light curves to obtain a refined period of 8.99$\pm$0.01~days, fully consistent with that found by \cite{walter2006} with {\em INTEGRAL} and similar to the 8.96$\pm$0.01~days found by \cite{corbet2005} with the first year of BAT data available at that date. We divided the full 12-year span into ten intervals and checked for the consistency of the period found. In Fig.~\ref{swiftbat} we show the best-period found using the {\sc efsearch} task of  {\sc heasoft} and the resulting folded X-ray light curve in the 15--50~keV band where a clear modulation of 0.6 to 1.8~cts~s$^{-1}$ was obtained for the full $\sim$4500-day light curve. The period reference corresponds to MJD 54702.82292, fixed at the middle of ObsID 0556140701 from the XMM-{\em Newton} campaign.

\section{Results}

\subsection{XMM-{\em Newton} light curves}
\label{split}

\begin{figure}
\centering
\includegraphics[angle=-90,width=\hsize]{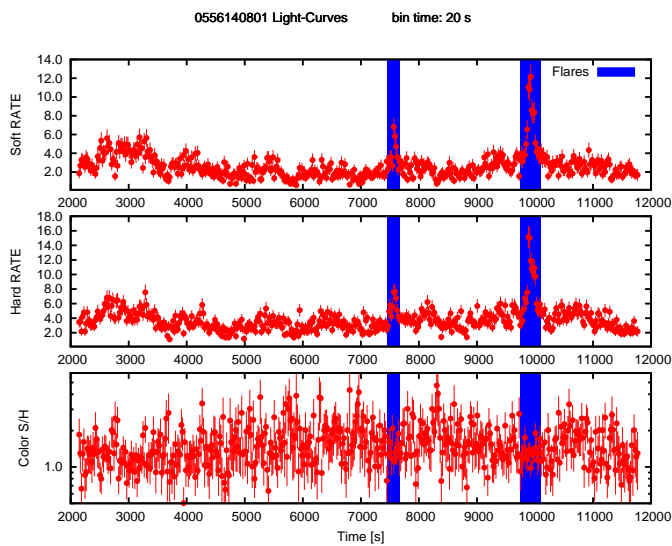}
   \caption{XMM-{\em Newton} PN background-corrected light curves of observation 0556140801  of IGR~J16320--4751 using 20~s binning time. {\bf Top panel:} soft band (0.5--6.0~keV) rate in cts~s$^{-1}$. {\bf Central panel:} hard band (6.0--12.0~keV) rate in cts~s$^{-1}$. {\bf Bottom panel:} soft-to-hard color ratio. Blue stripes correspond to two flaring intervals where the rate significantly increases in both soft and hard bands, keeping an average color value consistent with the rest of the observation. Error bars correspond to the 1-$\sigma$ confidence level.}
      \label{lc_0801}
\end{figure}

\begin{figure}
\centering
\includegraphics[angle=-90,width=\hsize]{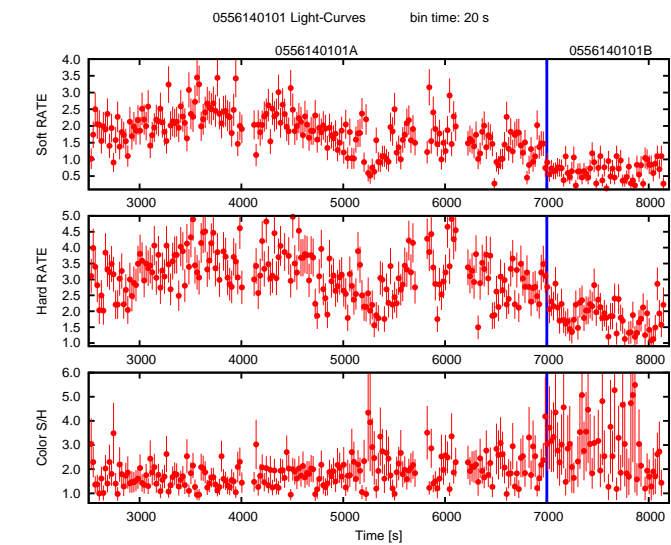}
   \caption{XMM-{\em Newton} PN background-corrected light curves of observation 0556140101  of IGR~J16320--4751 using 20~s binning time. {\bf Top panel:} soft band (0.5--6.0~keV) rate in cts~s$^{-1}$. {\bf Central panel:} hard band (6.0--12.0~keV) rate in cts~s$^{-1}$. {\bf Bottom panel:} soft/hard color ratio. The blue lines indicate a significant variation in the source state at $\sim$7000~s both in the count-rates and in the color ratio. Error bars correspond to 1-$\sigma$ confidence level.}
      \label{lc_0101}
\end{figure}

\begin{figure}
\centering
\includegraphics[angle=-90,width=\hsize]{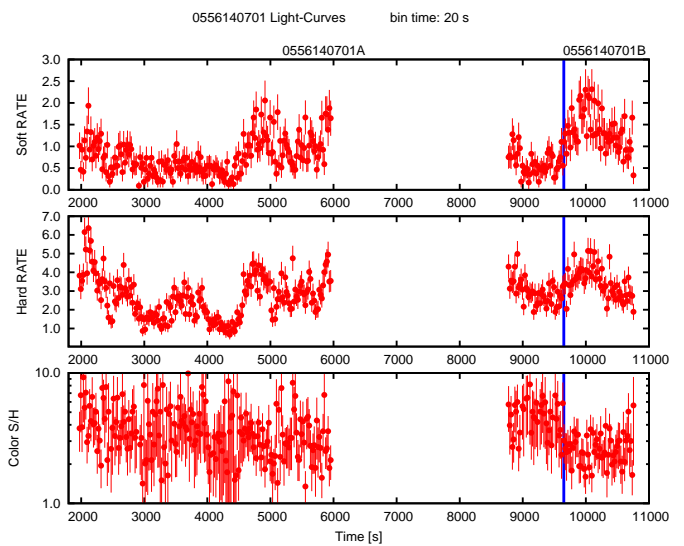}
   \caption{Same as Fig.~\ref{lc_0101}, but for  observation 0556140701.}
      \label{lc_0701}
\end{figure}

We extracted barycenter-corrected light curves using the annular regions determined by the pile-up analysis with an outer radius of 35''. For background regions we selected circles of 50'' in the same CCD of the source as suggested in the XMM-{\em Newton} Calibration Notes\footnote{http://xmm2.esac.esa.int/docs/documents/CAL-TN-0018.pdf}. Background-subtracted and exposure-corrected light curves were calculated using the {\sc epiclccorr} SAS task. 

Considering a total average count rate of 3.56~cts~s$^{-1}$, we chose a binning time of 20~s in order to analyze the X-ray variability and flaring activity of the source. Moreover, considering the overall shape of the spectra, we decided to use two energy bands to analyze the source color evolution: a {\it soft} band in the 0.5--6.0~keV energy range and a {\it hard} band in the 6.0--12.0~keV energy range. Then, by means of the {\sc lcurve} task from {\sc heasoft} we created {\it soft}, {\it hard}, and {\it soft/hard} ratio light curves that we used to search for long-term variability, short flaring, and possible spectral changes of the source.

\begin{figure*}
\centering
\includegraphics[height=0.49\hsize,angle=-90]{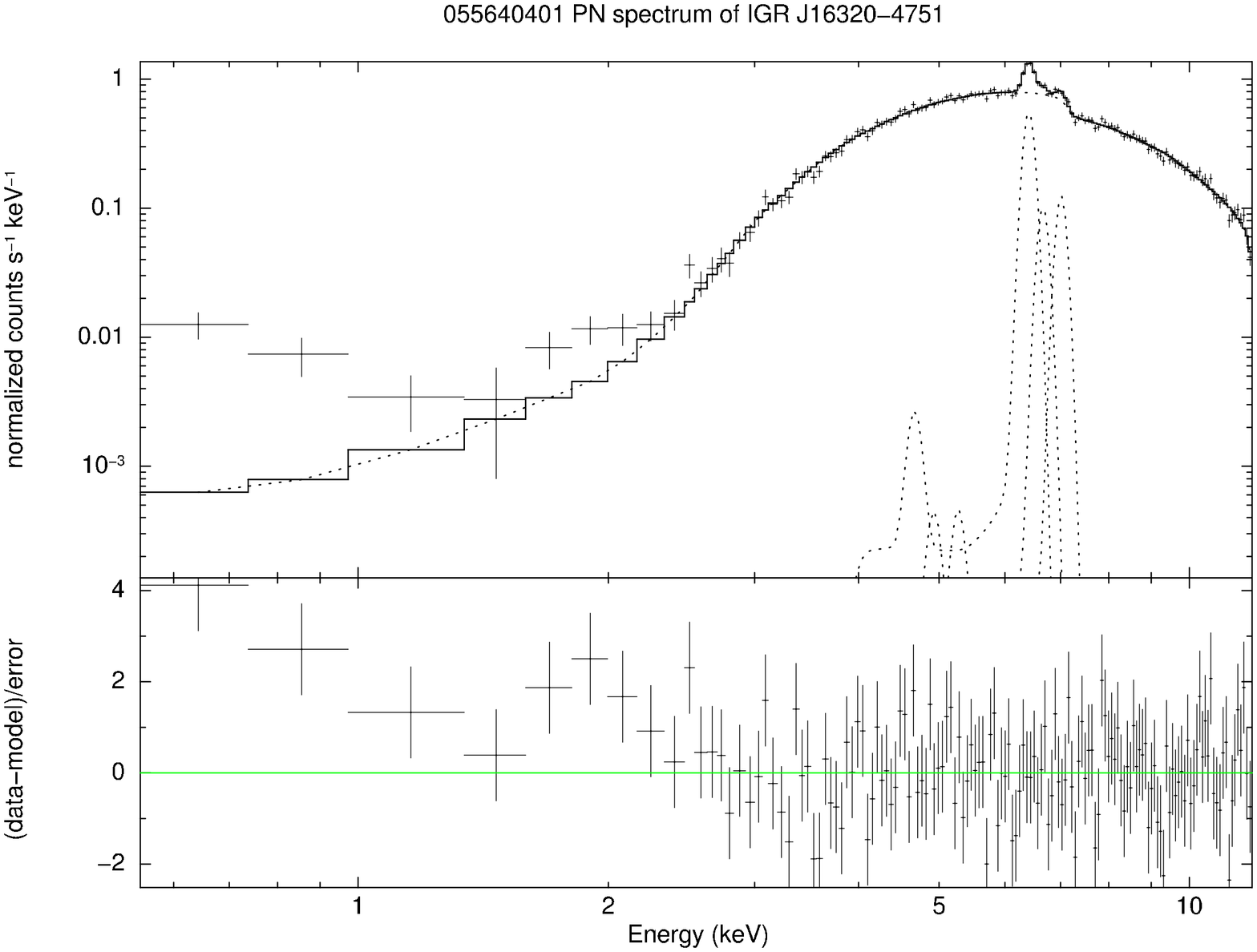}  
\includegraphics[height=0.49\hsize,angle=-90]{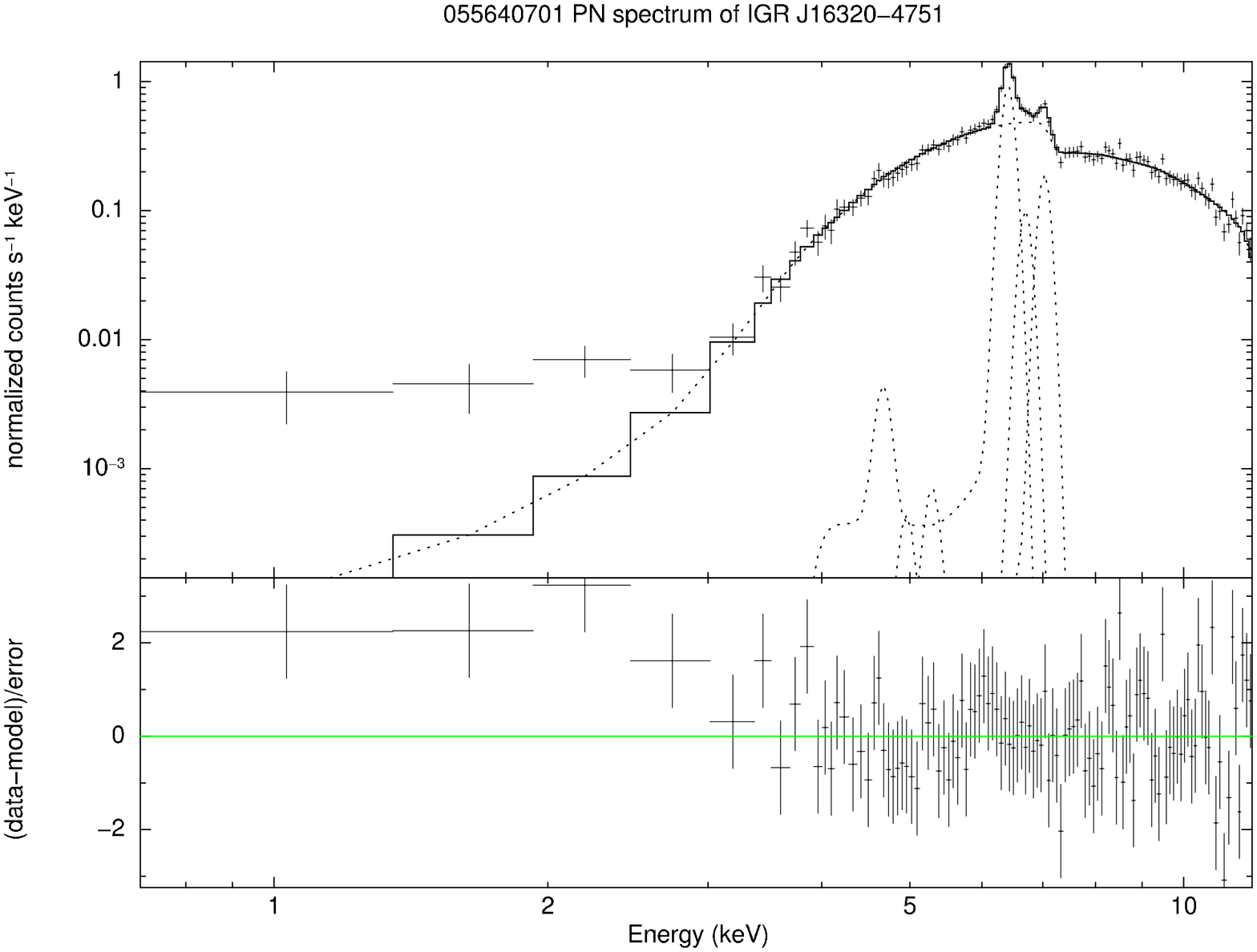}
   \caption{XMM-{\em Newton} PN spectra and best-fit model of IGR~J16320--4751 in two extreme cases. {\bf Left panel: } Background-subtracted X-ray spectra of ObsID 401. {\bf Right panel: } Background-subtracted X-ray spectra of ObsID 701. Lower panels show the fitting residuals. Errors are at 1-$\sigma$ (68\%) confidence levels, and $\chi^{2}$ statistics are used.}
      \label{plotspectrum}
\end{figure*}

Our visual inspection of the light curves indicated high variability on several timescales, including an overall change that can be seen on the variable net count rate in Table~\ref{obstable}. In some of them a clear modulation can be seen on a timescale of $\sim$1300~s, as was first pointed out by \cite{lutovinov2005}, a period that was attributed to the spin of the NS in the system. This modulation is evident in ObsIDs 0128531101, 0201700301, 0556140401, and 0556140601, both in the {\it soft} and {\it hard} bands, keeping the color ratio constant. This modulation persists in the rest of the observations, but it is not that clear. ObsIDs 0556140401 and 0556140601 show a secular increase in their count rates, evidencing the presence of a long-term variability of the source on a timescale of 10~ks. In ObsID 0556140801, two short and bright flares were detected when the source increases its rate by a factor of $\sim$10 for $\sim300$~s, keeping a constant color ratio (see Fig.~\ref{lc_0801}). For subsequent spectral analysis, we thus split this observation into 0556140801NF for the non-flaring intervals and 0556140801F for the flaring ones (indicated with blue bands in Fig.~\ref{lc_0801}). Even though the color ratio is $\sim$1.2--1.4 through all the observations, two strong variations can be seen in ObsIDs 0556140101 and 0556140701 where the ratio changes from $\sim$1.5 to $\sim$3 in the first case and from $\sim$4 to $\sim$2 in the second. We decided to split these observations, naming them  A and B as indicated by the blue  lines in Figs.~\ref{lc_0101}~and~\ref{lc_0701}. In Table~\ref{obstable} we present the average count rates and colors for each XMM-{\em Newton} observation.

The similar behavior observed in both the soft and hard X-ray light curves is the reason why the hardness ratio does not show significant variations between the flaring and non-flaring periods. This indicates that flares are related to a broadband flux increase and not to variations in the absorption. In this case, the hard band should be much less affected than the soft band, thus implying changes in the hardness evolution.

\subsection{XMM-{\em Newton} spectra}

We extracted source and background spectra from the same regions indicated in the light curve analysis. Redistribution response matrices (RMF) and ancillary response files (ARF) were generated using {\sc rmfgen} and {\sc arfgen} tasks, respectively, considering the {\it psf} function. We binned the spectra to obtain at least 25 counts per bin and we fitted them with {\sc XSPEC} v12.9.1 \citep{arnaud1996} considering the full 0.5--12.0~keV energy band.

\renewcommand{\arraystretch}{1.25}

\begin{table*}
\caption{Best-fit spectral parameters for the continuum emission.  Intrinsic absorption column density, $N_{\rm H}$, is indicated in units of 10$^{22}$~cm$^{-2}$. Plasma and electron temperatures, $kT_0$ and $kT_{\rm e}$, are shown in eV and keV, respectively. Optical depth, $\tau$, and normalization, Norm, correspond to {\sc comptt} model. Flux$_{\rm T}$ (0.5--12~keV), Flux$_{\rm S}$ (0.5--6~keV), and Flux$_{\rm H}$ (6--12~keV) are the unabsorbed X-ray fluxes in 10$^{-11}$~erg~cm$^{-2}$~s$^{-1}$.}
\label{spectra_continuum}
\centering
\setlength{\tabcolsep}{5pt}
\begin{tabular}{lclcccccccc} 
\hline\hline             
ObsID & Phase & Name & $N_{\rm H}$ & $kT_0$ & $kT_{\rm e}$ & $\tau$ & Norm & Flux$_{\rm T}$  & Flux$_{\rm S}$ & Flux$_{\rm H}$ \\
\hline
0128531101      &       $-$     &       1285    &       $       22.4    _{      -1.1    }^{+    0.9     }$ &       $       < 26.7$ &       $       2.56    _{      -0.11   }^{+    0.13    }$ &       $       14.1    _{      -1.8    }^{+    1.7     }$ &    $       7.7     _{      -1.7    }^{+    3.1     }$ &       $       2.8     _{      -0.6    }^{+    0.7     }$ &    $       1.3     _{      -0.4    }^{+    0.6     }$ &       $       1.5     _{      -0.1    }^{+    0.1     }$ \\
0201700301      &       $-$     &       2017    &       $       18.9    _{      -0.6    }^{+    0.1     }$ & $     <76.3$ &        $       2.77    _{      -0.06   }^{+    0.03    }$ &       $       17.9    _{      -0.4    }^{+    1.6     }$ &    $       16.1    _{      -2.1    }^{+    0.3     }$ &               $       13.4    _{      -0.5    }^{+    0.5     }$ &    $       5.2     _{      -0.4    }^{+    0.4     }$ &       $       8.2     _{      -0.1    }^{+    0.1     }$ \\
0556140101A     &       $       0.407   \pm     0.005   $       &       101A    &       $       37.6    _{      -1.1    }^{+    0.9     }$ &       $       71.0    _{      -2.8    }^{+    2.9     }$ &    $       3.47    _{      -0.22   }^{+    0.27    }$ &       $       14.1    _{      -1.7    }^{+    2.8     }$ &    $       13.0    _{      -1.5    }^{+    1.1     }$ &               $       20.3    _{      -1.6    }^{+    1.9     }$ &    $       7.6     _{      -1.3    }^{+    1.5     }$ &       $       12.7    _{      -0.4    }^{+    0.4     }$      \\
0556140101B     &       $       0.412   \pm     0.001   $       &       101B    &       $       47.2    _{      -2.0    }^{+    5.6     }$ &       $       71.1    _{      -16.7   }^{+    6.0     }$ &    $       2.96    _{      -0.16   }^{+    0.27    }$ &       $       100.1   _{      -64.9   }^{+    22.0    }$ &    $       4.1     _{      -0.2    }^{+    1.0     }$ &               $       9.6     _{      -0.7    }^{+    0.3     }$ &    $       2.5     _{      -0.4    }^{+    1.3     }$ &       $       7.2     _{      -0.5    }^{+    0.6     }$      \\
0556140201      &       $       0.606   \pm     0.005   $       &       201     &       $       26.8    _{      -0.6    }^{+    1.1     }$ &       $       71.0    _{      -3.5    }^{+    6.2     }$ &    $       4.91    _{      -0.07   }^{+    0.17    }$ &       $       12.4    _{      -0.5    }^{+    0.3     }$ &    $       18.8    _{      -0.4    }^{+    0.6     }$ &               $       35.7    _{      -2.0    }^{+    2.1     }$ &    $       12.4    _{      -1.6    }^{+    1.5     }$ &       $       23.5    _{      -0.6    }^{+    0.6     }$      \\
0556140301      &       $       0.809   \pm     0.005   $       &       301     &       $       21.6    _{      -1.5    }^{+    3.1     }$ &       $       71.0    _{      -10.8   }^{+    1.5     }$ &    $       2.85    _{      -0.12   }^{+    0.14    }$ &       $       35.9    _{      -12.1   }^{+    7.8     }$ &    $       7.6     _{      -0.5    }^{+    1.8     }$ &               $       15.5    _{      -1.1    }^{+    2.7     }$ &    $       4.5     _{      -1.1    }^{+    2.2     }$ &       $       10.9    _{      -0.5    }^{+    0.7     }$      \\
0556140401      &       $       0.006   \pm     0.007   $       &       401     &       $       29.3    _{      -0.4    }^{+    0.5     }$ &       $       71.0    _{      -0.5    }^{+    0.9     }$ &    $       3.33    _{      -0.05   }^{+    0.08    }$ &       $       15.7    _{      -0.6    }^{+    0.4     }$ &    $       13.3    _{      -0.2    }^{+    0.4     }$ &               $       21.7    _{      -0.9    }^{+    1.0     }$ &    $       7.8     _{      -0.8    }^{+    0.8     }$ &       $       14.0    _{      -0.2    }^{+    0.2     }$      \\
0556140501      &       $       0.109   \pm     0.001   $       &       501     &       $       31.7    _{      -2.7    }^{+    5.4     }$ &       $       71.0    _{      -8.3    }^{+    56.1    }$ &    $       3.15    _{      -0.22   }^{+    0.84    }$ &       $       14.2    _{      -5.0    }^{+    3.9     }$ &    $       12.2    _{      -2.0    }^{+    2.6     }$ &               $       17.3    _{      -3.3    }^{+    3.5     }$ &    $       6.9     _{      -2.6    }^{+    2.9     }$ &       $       10.4    _{      -0.8    }^{+    0.7     }$      \\
0556140601      &       $       0.212   \pm     0.008   $       &       601     &       $       25.8    _{      -0.3    }^{+    0.3     }$ &       $       70.9    _{      -0.1    }^{+    0.1     }$ &    $       3.09    _{      -0.04   }^{+    0.05    }$ &       $       20.3    _{      -0.5    }^{+    0.5     }$ &    $       24.7    _{      -0.2    }^{+    0.1     }$ &               $       43.8    _{      -1.6    }^{+    2.3     }$ &    $       14.5    _{      -1.4    }^{+    1.6     }$ &       $       29.6    _{      -0.5    }^{+    0.5     }$      \\
0556140701A     &       $       0.499   \pm     0.005   $       &       701A    &       $       58.6    _{      -1.2    }^{+    1.8     }$ &       $       70.9    _{      -4.0    }^{+    3.6     }$ &    $       3.84    _{      -0.05   }^{+    0.05    }$ &       $       39.0    _{      -1.3    }^{+    0.7     }$ &    $       6.8     _{      -0.1    }^{+    0.2     }$ &               $       16.8    _{      -1.1    }^{+    1.7     }$ &    $       3.6     _{      -0.5    }^{+    1.1     }$ &       $       13.2    _{      -0.5    }^{+    0.5     }$      \\
0556140701B     &       $       0.505   \pm     0.001   $       &       701B    &       $       58.7    _{      -5.8    }^{+    4.0     }$ &       $       71.0    _{      -4.2    }^{+    21.0    }$ &    $       7.44    _{      -1.62   }^{+    0.63    }$ &       $       9.5     _{      -0.3    }^{+    2.0     }$ &    $       13.9    _{      -2.0    }^{+    2.6     }$ &               $       28.6    _{      -4.1    }^{+    3.7     }$ &    $       10.8    _{      -2.9    }^{+    2.6     }$ &       $       17.8    _{      -1.2    }^{+    1.2     }$      \\
0556140801F     &       $       0.700   \pm     0.001   $       &       801F    &       $       30.2    _{      -2.0    }^{+    3.3     }$ &       $       301.4   _{      -205.3  }^{+    416.6   }$ &    $       3.88    _{      -0.02   }^{+    0.49    }$ &       $       10.4    _{      -1.9    }^{+    1.5     }$ &    $       20.2    _{      -3.9    }^{+    2.5     }$ &               $       38.5    _{      -6.9    }^{+    6.6     }$ &    $       16.5    _{      -5.4    }^{+    5.5     }$ &       $       22.0    _{      -1.7    }^{+    1.5     }$      \\
0556140801NF    &       $       0.700   \pm     0.006   $       &       801NF   &       $       22.9    _{      -0.3    }^{+    0.6     }$ &       $       71.0    _{      -2.0    }^{+    0.9     }$ &    $       2.72    _{      -0.03   }^{+    0.03    }$ &       $       31.8    _{      -1.4    }^{+    1.1     }$ &    $       8.6     _{      -0.1    }^{+    0.2     }$ &               $       16.3    _{      -1.0    }^{+    1.4     }$ &    $       5.0     _{      -0.5    }^{+    1.2     }$ &       $       11.1    _{      -0.3    }^{+    0.1     }$      \\
0556141001      &       $       0.090   \pm     0.005   $       &       1001    &       $       25.2    _{      -1.0    }^{+    1.0     }$ &       $       71.0    _{      -1.9    }^{+    1.6     }$ &    $       3.52    _{      -0.14   }^{+    0.17    }$ &       $       16.1    _{      -1.5    }^{+    2.1     }$ &    $       13.6    _{      -1.0    }^{+    0.8     }$ &               $       23.9    _{      -1.3    }^{+    1.4     }$ &    $       8.2     _{      -1.1    }^{+    1.0     }$ &       $       15.8    _{      -0.4    }^{+    0.3     }$      \\
\hline
\end{tabular}
\end{table*}

\renewcommand{\arraystretch}{1.25}
\begin{table*}
\caption{Best-fit spectral parameters of the Gaussian emission lines. $E_{\rm K\alpha}$, $E_{\rm XXV}$, and $E_{\rm K\beta}$ are the central energies of the lines expressed in keV and Flux$_{\rm K\alpha}$, Flux$_{\rm XXV}$, and Flux$_{\rm K\beta}$ are their corresponding fluxes in units of 10$^5$~photons~cm$^{-2}$~s$^{-1}$. The $\dagger$ symbol indicates parameters that were fixed during the fits.}
\label{spectra_lines}
\centering
\begin{tabular}{lccccccc} 
\hline\hline  
ObsID & $E_{\rm K\alpha}$ & Flux$_{\rm K\alpha}$ & $E_{\rm XXV}$ & Flux$_{\rm XXV}$ & $E_{\rm K\beta}$ & Flux$_{\rm K\beta}$  & $\chi^2_\nu$ / d.o.f.\\
\hline  
0128531101 & 
$       6.40    _{      -0.05   }^{+    0.16    }$ &    $       1.5     _{      -0.7    }^{+    0.7     }$ &       $       6.68    ^{              \dagger         }$ &    $       0.00    ^{              \dagger         }$ &       $       7.19    _{      -0.24   }^{+    0.04    }$ &    $       0.96    _{      0.00    }^{+    0.14    }$ &       0.84    /       70      \\
0201700301 & 
$       6.42    _{      -0.01   }^{+    0.01    }$ &    $       17.7    _{      -0.6    }^{+    0.2     }$ &       $       6.72    _{      -0.05   }^{+    0.06    }$ &    $       3.59    _{      -0.08   }^{+    0.03    }$ &       $       7.06    _{      -0.06   }^{+    0.05    }$ &    $       3.80    _{      -0.03   }^{+    0.07    }$ &       1.17    /       150     \\
0556140101A &
$       6.42    _{      -0.01   }^{+    0.01    }$ &    $       48.1    _{      -3.2    }^{+    3.4     }$ &       $       6.68    ^{              \dagger         }$ &    $       2.96    _{      -0.31   }^{+    0.39    }$ &       $       7.04    _{      -0.05   }^{+    0.04    }$ &    $       6.84    _{      -1.07   }^{+    0.07    }$ &       1.03    /       108     \\
0556140101B &
$       6.41    _{      -0.03   }^{+    0.02    }$ &    $       51.1    _{      -7.6    }^{+    5.4     }$ &       $       6.68    ^{              \dagger         }$ &    $       0.00    ^{              \dagger         }$ &       $       7.05    _{      -0.06   }^{+    0.07    }$ &    $       11.93   _{      -4.15   }^{+    3.76    }$ &       0.51    /       38      \\
0556140201 &
$       6.42    _{      -0.01   }^{+    0.01    }$ &    $       66.0    _{      -4.7    }^{+    4.1     }$ &       $       6.73    _{      -0.04   }^{+    0.07    }$ &    $       10.30   _{      -1.69   }^{+    0.65    }$ &       $       7.01    _{      -0.05   }^{+    0.06    }$ &    $       14.23   _{      -3.63   }^{+    0.92    }$ &       1.08    /       112     \\
0556140301 &
$       6.41    _{      -0.03   }^{+    0.03    }$ &    $       32.7    _{      -5.2    }^{+    3.8     }$ &       $       6.68    ^{              \dagger         }$ &    $       0.00    ^{              \dagger         }$ &       $       6.97    _{      -0.04   }^{+    0.11    }$ &    $       7.90    _{      -2.68   }^{+    1.79    }$ &       1.11    /       78      \\
0556140401 &
$       6.41    _{      -0.01   }^{+    0.01    }$ &    $       35.6    _{      -1.5    }^{+    1.3     }$ &       $       6.66    _{      -0.03   }^{+    0.03    }$ &    $       6.66    _{      -0.07   }^{+    0.05    }$ &       $       7.01    _{      -0.03   }^{+    0.04    }$ &    $       8.48    _{      -0.30   }^{+    0.30    }$ &       1.22    /       134     \\
0556140501 &
$       6.43    _{      -0.05   }^{+    0.02    }$ &    $       23.8    _{      -6.4    }^{+    8.6     }$ &       $       6.68    ^{              \dagger         }$ &    $       0.008   _{      -0.003  }^{+    0.003   }$ &       $       6.90    _{      0.01    }^{+    0.26    }$ &    $       9.16    _{      -0.09   }^{+    3.12    }$ &       0.77    /       66      \\
0556140601 &
$       6.41    _{      -0.01   }^{+    0.01    }$ &    $       101.7   _{      -0.9    }^{+    0.5     }$ &       $       6.73    _{      -0.04   }^{+    0.02    }$ &    $       14.26   _{      -0.31   }^{+    0.26    }$ &       $       7.05    _{      -0.01   }^{+    0.01    }$ &    $       30.88   _{      -0.91   }^{+    0.88    }$ &       1.08    /       140     \\
0556140701A &
$       6.41    _{      -0.01   }^{+    0.01    }$ &    $       81.9    _{      -3.6    }^{+    3.4     }$ &       $       6.68    _{      -0.04   }^{+    0.03    }$ &    $       10.74   _{      -0.28   }^{+    0.09    }$ &       $       7.01    _{      -0.02   }^{+    0.02    }$ &    $       21.08   _{      -0.45   }^{+    0.30    }$ &       1.00    /       94      \\
0556140701B &
$       6.39    _{      -0.01   }^{+    0.03    }$ &    $       85.8    _{      -11.2   }^{+    11.1    }$ &       $       6.75    _{      -0.12   }^{+    0.04    }$ &    $       11.29   _{      -0.71   }^{+    3.37    }$ &       $       7.02    ^{              \dagger         }$ &    $       1.77    _{      -0.55   }^{+    1.35    }$ &       1.07    /       66      \\
0556140801F &
$       6.44    _{      -0.01   }^{+    0.06    }$ &    $       68.2    _{      -3.2    }^{+    7.8     }$ &       $       6.68    ^{              \dagger         }$ &    $       0.00    ^{              \dagger         }$ &       $       7.02    ^{              \dagger         }$ &    $       0.00    ^{              \dagger         }$ &       1.00    /       57      \\
0556140801NF &
$       6.40    _{      -0.01   }^{+    0.01    }$ &    $       38.6    _{      -1.5    }^{+    1.8     }$ &       $       6.63    _{      -0.02   }^{+    0.06    }$ &    $       9.69    _{      -0.16   }^{+    0.14    }$ &       $       6.97    _{      -0.03   }^{+    0.03    }$ &    $       11.21   _{      -0.40   }^{+    0.41    }$ &       1.28    /       119     \\
0556141001 &
$       6.41    _{      -0.02   }^{+    0.01    }$ &    $       36.8    _{      -0.8    }^{+    1.3     }$ &       $       6.68    _{      -0.04   }^{+    0.10    }$ &    $       4.74    _{      -0.09   }^{+    0.27    }$ &       $       7.03    _{      -0.04   }^{+    0.07    }$ &    $       9.47    _{      -0.18   }^{+    0.14    }$ &       1.30    /       127     \\
\hline
\end{tabular}
\end{table*}

The spectra of IGR~J16320--4751 are characterized by a highly absorbed continuum 
and a clear Fe-edge of absorption at $\sim$7~keV, which varies significantly among the different observations. 
In ObsID 0556140401, only 0.4\% of the counts are in the 0.5--2.0~keV band, and this fraction decreases to 0.3\% and 0\% for ObsIDs 0556140701A and 0556140701B, respectively.
The continuum has a power law-like shape with a high-energy cutoff that becomes evident at energies above 7--9~keV. Prominent Fe-K$\alpha$ lines at $\sim$6.4~keV are present in all the spectra. In the best-quality spectra, fainter Fe-K$\beta$ (at $\sim$7~keV) and Fe XXV (at $\sim$6.7~keV) lines can also be seen. We fitted the spectra by means of a thermally Comptonized model {\sc comptt} \citep{titarchuk} for the continuum and three narrow Gaussian functions (with null widths $\sigma = 0$~keV) to model the emission lines. {\sc comptt} is an analytical model that describes the Comptonization of soft photons with temperature $kT_0$ by a hot plasma with temperature $kT_{\rm e}$ and optical depth $\tau$. We included two {\sc tbabs} absorption components using abundances from \cite{wilms} to attain for both the interstellar medium (ISM) and the intrinsic absorption of the obscured HMXB system. We fixed the hydrogen column density of the ISM absorption model to $2.1\times10^{22}$~cm$^{-2}$ as in \cite{rodriguez2003} and we let the second hydrogen column $N_{\rm H}$ to freely vary during the fits. 

We estimated confidence regions for all parameters at 90\% level using the Markov chain Monte Carlo (MCMC) method implemented in XSPEC. For our calculations we ran eight walkers for a total of $8 \times 10^4$ steps to find the best-fit values of the free parameters together with their confidence regions (see Tables~\ref{spectra_continuum}~and~\ref{spectra_lines})  as well as the reduced $\chi^2_\nu$ and the degrees of freedom (d.o.f.). We used the {\sc cflux} convolution model to estimate the unabsorbed flux of the continuum Comptonization in the {\em soft}, {\em hard}, and {\em total} (0.5--12.0~keV) bands. Shortened ObsIDs are shown in the  ``Name'' column. Phases correspond to an orbital period of 8.99~days and a central epoch corresponding to the middle of the 0556140701 exposure (phase=0.5). Phases are centered on each observation and their symmetrical error bars correspond to their respective duration. In Fig.~\ref{plotspectrum}, we show two examples of the XMM-{\em Newton} PN spectra of IGR~J16320--4751. In the left (right) panel we show background-subtracted X-ray spectra of ObsID 401 (701). Errors are at 1-$\sigma$ confidence levels, and $\chi^{2}$ statistics are used.

\begin{figure*}
\centering
\includegraphics[width=0.48\hsize]{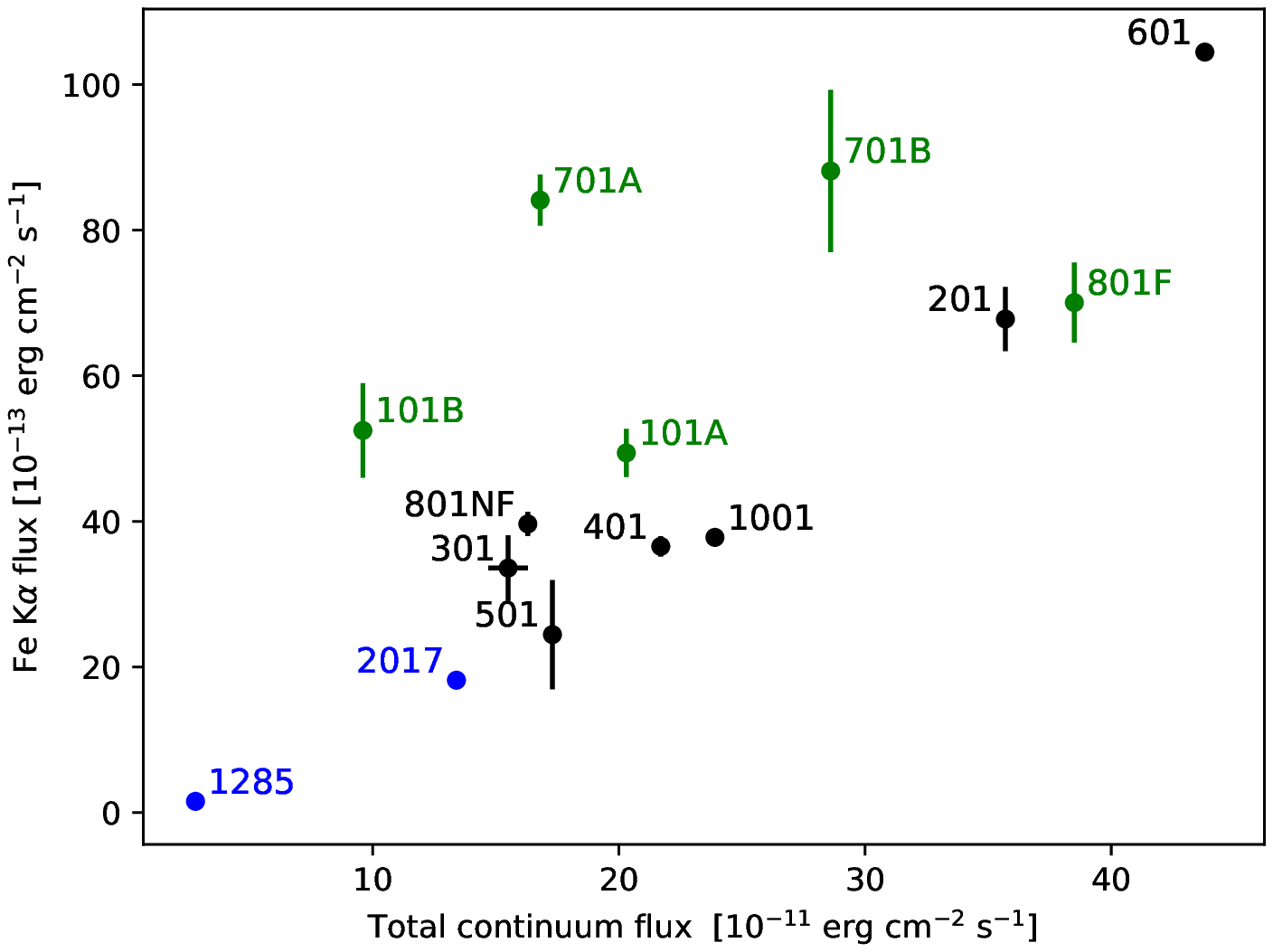}
\includegraphics[width=0.48\hsize]{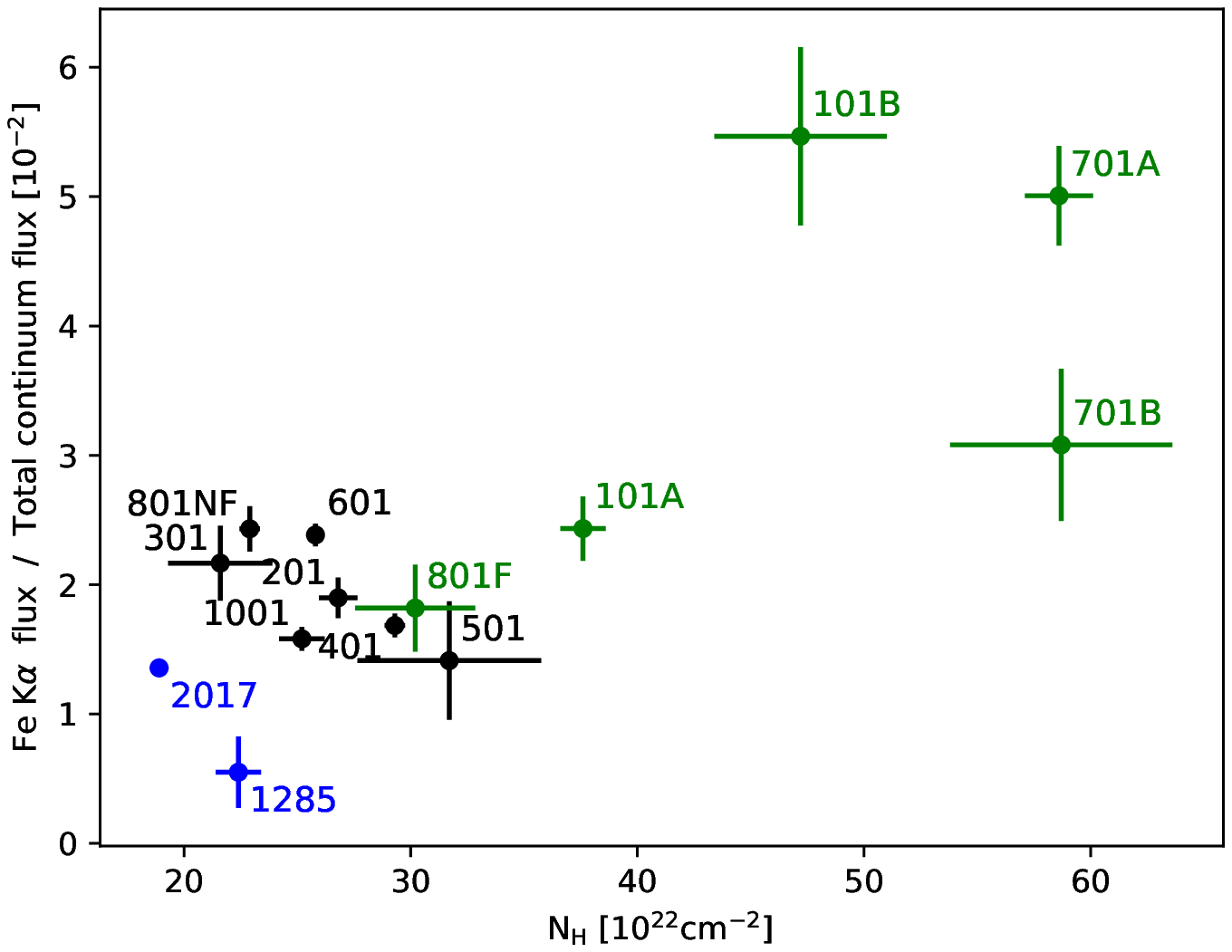}
   \caption{Correlations between spectral parameters of IGR~J16320--4751. {\bf Left panel:} Flux$_{\rm K\alpha}$ with respect to Flux$_{\rm T}$. {\bf Right panel:} Flux$_{\rm K\alpha}$ normalized to Flux$_{\rm T}$, with respect to the intrinsic $N_{\rm H}$. Error bars correspond to the 90\% confidence level.  Blue dots correspond to earlier observations from 2003 and 2004. Green dots indicate split observations (see Sect.~\ref{split}). The figures suggest two possible correlations between these spectral parameters for the whole set of XMM-{\em Newton} observations.  Using a Pearson test we found $R=0.76$ with a $p$-value = 0.002 for the left panel and $R=0.77$ for a probability $p$-value = 0.001 for the right panel. } 
      \label{NHFeKa}

\end{figure*}

The good quality of the fits obtained with the {\sc comptt} model suggests that the continuum emission is compatible with a Comptonization of soft photons, emitted close to the surface of the NS by a diluted cloud of hot electrons surrounding the compact object. However, as first noticed by \cite{rodriguez2003}, the lack of data above 12~keV, does not allow us to obtain a good constraint on the cutoff energy. Regarding the intrinsic absorption column density, two important facts should be noted. First, in the majority of the observations, we obtain $N_{\rm H} \sim 20-30 \times 10^{22}$ cm$^{-2}$, which is fully compatible with the values found by \cite{rodriguez2006} with XMM-{\em Newton} and \cite{intzand2003} in a BeppoSAX spectrum. Second, in Obs 101A, 101B, 701A, and 701B, we found significantly higher values of N$_{\rm H} \sim 35-60 \times 10^{22}$ cm$^{-2}$ without noticing strong changes in the continuum emission parameters, which suggests that this variation is a geometrical effect instead of a local sudden change in the accretion process.

Despite the good quality of the fits, a systematic trend to a soft excess is suggested by the spectra (see Fig.~\ref{plotspectrum}). In order to test for possible systematics that could make us overpredict the $N_{\rm H}$ column value, we did two different tests. First, we added a blackbody to account for a possible soft-excess origin and fitted  the spectra again, re-calculating the 90\% confidence region for the $N_{\rm H}$ value. For ObsID 0401, we obtained $(28-30.7) \times 10^{22}$~cm$^{-2}$, which is fully consistent with the value reported in Table~\ref{spectra_continuum}. For ObsID 0701A, we found $(57.9-71.7) \times 10^{22}$~cm$^{-2}$, which is consistent in the lower bound and increases for the higher bound, due to a correlation between the $N_{\rm H}$ and the blackbody. As a second test, we fitted the same spectra restricting ourselves to the 2--12~keV energy band. For 0701A, we found that the $N_{\rm H}$ column is within the $(54.3-60.9) \times 10^{22}$~cm$^{-2}$ range, while for 0401 it remains in the $(28.4-30.5) \times 10^{22}$~cm$^{-2}$ range, being highly consistent with the values presented in Table~\ref{spectra_continuum} for the full 0.5--12~keV spectral range.

Following the conversion expression $N_{\rm H}$~[cm$^{-2}$] = $(2.21 \pm 0.09) \times 10^{21}$~$A_{\rm V}$~[mag] \citep{guver2009}, and considering the minimum X-ray column of $N_{\rm H} = 2.1 \times 10^{23}$~cm$^{-2}$ found in our series of spectra, we deduce an optical extinction of $A_{\rm V} \sim 100$~mag, which is much higher than the optical extinction derived in the optical and the infrared bands \citep{chaty2008,coleiro2013}. In the above-mentioned papers, the authors report corresponding absorption columns of 2.14 and 6.60 $\times 10^{20}$~cm$^{-2}$ in the optical and IR bands, respectively. This difference in the absorption columns can be explained by stratification, where the observed radiation at different wavelengths  originates at different depths in the system, and thus transverses different optical paths, probing different emission and absorption layers. While in the optical/IR band we measure the absorption of optically thick material down to the surface of the supergiant star, in the X-ray band we measure the intrinsic absorption down to the surface of the compact object, which seems to be embedded in a dense cloud of accreting material \citep[similar to the case of IGR J16318-4848,][]{chatyrahoui2012}.

While the Fe~K$\alpha$ and Fe~K$\beta$ lines are clearly noticeable in all the X-ray spectra, Fe~XXV is not that prominent, being hardly detectable in some of the observations. In the cases where the spectral fits were not sensitive to the addition of this third Gaussian (e.g., 101B, 301, 801F), we decided to fix its normalization to zero during the fits (indicated by  $\dagger$  in Table~\ref{spectra_lines}). It is important to note that in all cases we were able to obtain very good fits to the emission line by means of zero-width Gaussians, meaning that the broadening of the Fe lines (when present) should be lower than the width of the response of the PN camera at those energies.

As previously noted by \cite{gimenezgarcia2015}, we also found a correlation between the flux of the Fe~K$\alpha$ line and the continuum level, characterized by the total unabsorbed flux in the 0.5--12.0~keV energy band (see  left panel of Fig.~\ref{NHFeKa}). This is expected for fluorescence emission from a small region of matter in the very close vicinity of the illuminating compact object. We also confirm a correlation between the X-ray luminosity and the Fe K$\alpha$ flux, normalized by the X-ray flux, known as Baldwin effect. As pointed out by \cite{gimenezgarcia2015}, $\gamma$ Cas sources do not follow this correlation, suggesting a different emission scenario in sgHMXBs like IGR~J16320--4751.

\section{Discussion}

\begin{figure}
\centering
\includegraphics[width=\hsize]{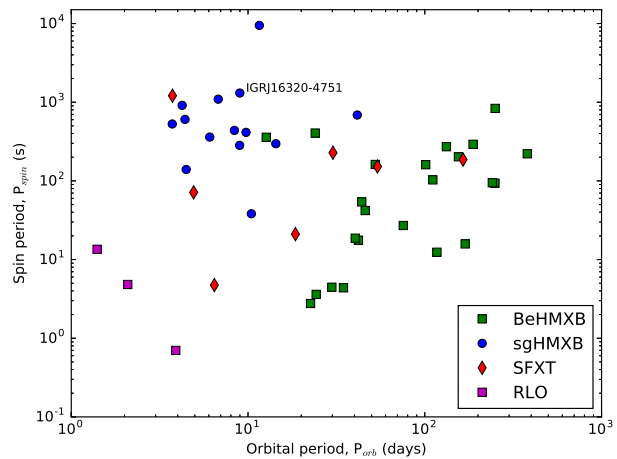}
   \caption{Updated Corbet diagram (1986) showing the different populations of HMXBs (X-ray pulsars) with measured values of both $P_{\rm orb}$ and $P_{\rm spin}$. BeHMXB show a correlation between $P_{\rm orb}$ and $P_{\rm spin}$ due to net transfer of angular momentum between the decretion disk of the Be companion star and the neutron star. Beginning atmospheric Roche-lobe overflow (RLO) systems have shorter (likely circularized) orbital periods. Supergiant fast X-ray transients (SFXT) are accreting pulsars exhibiting short and intense flares, spanning nearly the whole range of parameters of this diagram. Finally, supergiant HMXBs are accreting pulsars with quite short orbital periods (most being circularized) and slowly rotating neutron stars. IGR~J16320--4751, labeled in this diagram, is one of these persistently accreting systems.}
      \label{corbet}
\end{figure}

In Fig.~\ref{corbet} we show the up-to-date Corbet diagram \citep{corbet1986} including all the well-known HMXBs hosting NSs. With an orbital period of $\sim$9~days, and a quite slowly rotating NS with spin period of $\sim$1300~s, IGR~J16320--4751 is a prototype of the sgHMXB class, as shown by the pulse-orbital period diagram. The measured X-ray column density, $N_{\rm H} \approx 2 \times 10^{23}$~cm$^{-2}$, is an order of magnitude higher than the absorption column density along the line of sight, favoring the fact that this source belongs to the intrinsically absorbed sgHMXB class.

In sgHMXBs like IGR~J16320--4751, the observed X-ray luminosity is powered by wind-accretion onto the NS. 
The compact object is usually embedded in a dense and powerful wind provided by the OB supergiant companion. The X-ray emission properties directly depend on both the wind geometry (velocity and density profile, inhomogeneities, etc.) and the orbit (semiaxis and eccentricity) which together determine the wind accretion rate. Supergiant stars studies show that these extended stellar winds are quite complex. 
The wind has a high density-gradient decreasing outwards from the star \citep[CAK model,][]{cakmodel}, and local random inhomogeneities or clumps are expected as an intrisic feature of the radiatively driven winds \citep{oskinova2012}. While the latter can be responsible for short flares like those found in ObsID 801, the orbital modulation of the X-ray spectra and the hard X-ray light curve could be explained by a simple geometrical model taking into account the wind density profile and the orbital geometry with respect to a distant observer. 

\begin{figure}
\centering
\includegraphics[trim={0cm 0cm 0cm 0cm},clip, width=0.6\hsize]{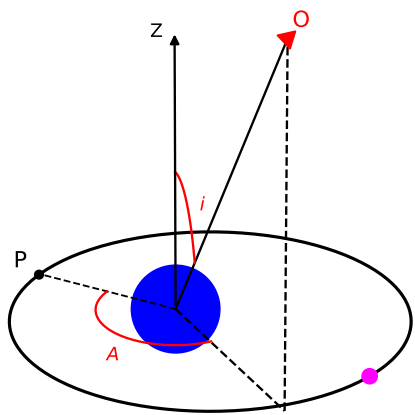}
   \caption{Schematic view of our model.  The neutron star (purple circle) describes an elliptical orbit in the XY-plane around the supergiant star (blue circle). A distant observer $O$ (red triangle) is located in the direction defined by its angular coordinates $i$ (inclination) and $A$ (azimuth) with respect to the orbital plane and periastron position $P$.}
      \label{modelito}
\end{figure}

With the aim of better understanding the orbital modulations found in the intrinsic absorption column density and the BAT hard X-ray light curve we propose a simple model consisting of a NS orbiting an OB supergiant embedded in its dense wind. Assuming a supergiant donor with $M_\star = 25$~M$_\odot$ and $R_\star = 20$~R$_\odot$ and a canonical NS of 1.4~$M_\odot$, for a $P = 8.99~d$, using  Kepler's third law, we obtain a semimajor axis of $a = 0.25$~AU which represents $\sim$2.7~$R_\star$. We model the supergiant wind profile by means of a typical CAK model with a mass-loss rate of $3 \times 10^{-6}$~M$_\odot$~yr$^{-1}$, $\beta = 0.85$ and $v_\infty = 1300$~km~s$^{-1}$. In such a close orbit, the matter captured by the NS as it moves along its elliptical orbit is able to produce a persistent X-ray emission exhibiting flux variability. This variability corresponds to the orbit eccentricity and to a periodical increase of the $N_{\rm H}$ column density when the NS is located close to the line connecting the observer and the supergiant. With this simple model, short flares cannot be explained; instead,  abrupt transitions in the accretion rate should correspond to the wind clumping. Once the stationary wind profile is defined, we still have three free parameters: the orbital eccentricity $e$, and the angular coordinates $A$ (azimuth) and $i$ (inclination) of a distant observer. These angles are measured with respect to the fixed orbital plane and periastron position $P$. A schematic view of our model can be seen in Fig.~\ref{modelito}.

\begin{figure}
\centering
\includegraphics[width=\hsize]{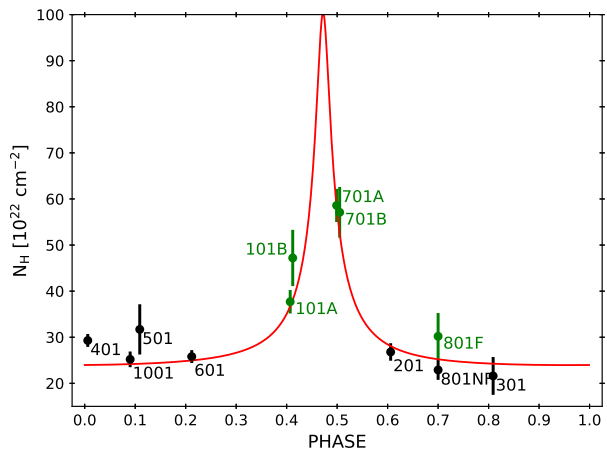}
\includegraphics[width=\hsize]{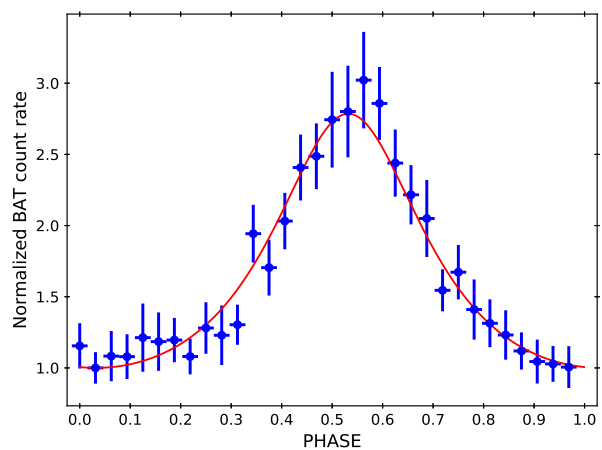}
   \caption{Orbital evolution of IGR~J16320--4751. {\bf Upper panel:} Intrinsic absorption column density, $N_{\rm H}$, obtained from XMM-{\em Newton} spectral fitting. Error bars correspond to the 90\% confidence level. {\bf Lower panel:} Folded {\em Swift}/BAT light curve normalized by its minimum value. Error bars correspond to the 1-$\sigma$ confidence level. Red lines are the absorption column density and normalized local density of our simple wind model scenario obtained for the parameters indicated in the text.}
      \label{NHBAT}
\end{figure}

In order to compute the $N_{\rm H}$ column density, for each orbital phase (or time), we integrated the wind density profile along the line connecting the observer $O$ with the NS position. Moreover, we assumed that the {\em Swift}/BAT count rate is proportional to the local wind density at the position of the accreting NS. Under all the above mentioned assumptions, we simultaneously fitted the {\em Swift}/BAT folded light curve and the $N_{\rm H}$ phase evolution obtaining an optimal solution given by $e=0.20 \pm 0.01$, $A=146.3^\circ$~$^{+3.7}_{-2.9}$, and $i=62.1^\circ$~$^{+0.3}_{-1.5}$ (90\% confidence intervals) that we present in Fig.~\ref{NHBAT}. The eccentricity is mainly constrained by the BAT light curve amplitude while the inclination depends strongly on the $N_{\rm H}$ increase profile. Our simple model is able to simultaneously fit both data sets, explaining the sudden change in the $N_{\rm H}$ column density, which peaks at a phase $\sim$0.47, and the BAT smooth modulation, which has its maximum at a phase $\sim$0.53. The phase difference between both maxima arises as a consequence of the observer azimuth angle.

The period uncertainty prevents us from accurately determining the phase of earlier XMM-{\em Newton} ObsIDs 0128531101 and 0201700301, and thus we decided to avoid including them in the analysis of the orbital evolution. However, for the sake of the correlation of different spectral parameters independent of their orbital phase, these observations were fully incorporated in our analysis.

Regarding the Fe~K$\alpha$ line, we were also able to recover a clear {curve of growth} as was previously found for other SGXBs sources, where the flux of the Gaussian normalized by the continuum flux is correlated with the $N_{\rm H}$ column (see right panel of Fig.~\ref{NHFeKa}). This  suggests the existence of a strong link between the absorbing and the fluorescent matter that, together with the orbital modulation, points to the stellar wind as the main contributor to both continuum absorption and Fe~K$\alpha$ emission in this source. Moreover, the narrowness of the  Fe lines  in all X-ray spectra allows us to confirm that IGR~J16320--4751 belongs to the HMXB narrow-line group \citep{gimenezgarcia2015}.

\section{Conclusions}

In this paper we reported a comprehensive work on IGR~J16320--4751, an archetype of the new class of highly absorbed HMXBs hosting supergiant stars, as shown by its localization in the Corbet diagram (Fig.~\ref{corbet}). After having refined the orbital period, we investigated the geometrical and physical properties of both components of the binary system along their full orbit by analyzing phase-resolved X-ray spectra obtained with XMM-{\em Newton} and the hard X-ray light curve from {\em Swift}/BAT. First, we found a clear modulation of both the intrinsic hydrogen column density and the hard X-ray light curve with the orbital phase (Fig.~\ref{NHBAT}). Second, we recovered two additional correlations:  one connects the Fe~K$\alpha$ line flux to the column density, suggesting that fluorescent matter is related to absorbing matter,  and the other  relates the Fe~K$\alpha$ line flux to the total continuum flux, suggesting that fluorescence emission emanates from a small region close to the accreting pulsar. Based on these two correlations we suggest that the absorbing matter is located within a small dense region surrounding the NS.

We then built a simple geometrical model of a NS orbiting a supergiant companion and accreting from its intense stellar wind. This simple model is able to reproduce the orbital modulation of both the sudden change in absorption column density and the smooth evolution of hard X-ray {\em Swift}/BAT folded light curve, as well as the phase shifts of their maxima. By putting together both correlations described above -- the Fe~K$\alpha$ line with column density and total continuum flux -- and these two successful fits -- the $N_{\rm H}$ and hard X-ray light curves -- we unambiguously show that the orbital modulation of the three observed parameters --column density, hard X-ray flux, and the Fe~K$\alpha$ line-- is caused by intrinsic absorption of matter surrounding the NS, modulated by the stellar wind density profile as viewed by the observer along the line of sight.

To conclude, this work provides strong support to our current understanding of intrinsically absorbed sgHMXBs. Additional studies applied to the spectral evolution analysis, and in particular time-resolved and polarimetric X-ray observations sampled along their full orbit, would be of high value to better constrain the overall physical and geometrical properties of sgHMXBs.

\begin{acknowledgements}
We made use of the IGR source webpage maintained by J. Rodriguez (http://irfu.cea.fr/Sap/IGR-Sources). We are grateful to Francis Fortin for insightful discussions. FG and SC were partly supported by the LabEx UnivEarthS, Interface project I10, ``From evolution of binaries to merging of compact objects.'' This work was partly supported by the Centre National d'Etudes Spatiales (CNES), and based on observations obtained with MINE: the Multi-wavelength {\em INTEGRAL} NEtwork. FG and JAC acknowledge support from PIP 0102 (CONICET). FAF is a fellow of CONICET. JAC is a CONICET researcher. This work received financial support from PICT-2017-2865 (ANPCyT). JAC was also supported on different aspects of this work by Consejer\'{\i}a de Econom\'{\i}a, Innovaci\'on, Ciencia y Empleo of Junta de Andaluc\'{\i}a under excellence grant FQM-1343 and research group FQM-322, as well as FEDER funds. 
\end{acknowledgements}


\begin{thebibliography}{aa}

\bibitem[Arnaud(1996)]{arnaud1996} Arnaud, K.~A.\ 1996, ASP Conf.~Ser.~101: Astronomical Data Analysis Software and Systems V, 101, 17.

\bibitem[Castor et al.(1975)]{cakmodel} Castor, J.~I., Abbott, D.~C., \& Klein, R.~I.\ 1975, \apj, 195, 157 

\bibitem[Chaty(2008)]{chaty2008} Chaty, S., Rahoui, F., Foellmi, C., Tomsick, J. A., Rodriguez, J., Walter, R.\ 2008, \aap, 484, 783

\bibitem[Chaty \& Rahoui(2012)]{chatyrahoui2012} Chaty, S., \& Rahoui, F.\ 2012, \apj, 751, 150 

\bibitem[Coleiro et al.(2013)]{coleiro2013} Coleiro, A., Chaty, S., Zurita Heras, J.~A., Rahoui, F., \& Tomsick, J.~A.\ 2013, \aap, 560, A108 

\bibitem[Corbet(1986)]{corbet1986} Corbet, R.~H.~D.\ 1986, \mnras, 220, 1047 

\bibitem[Corbet et al.(2005)]{corbet2005} Corbet, R., Barbier, L., Barthelmy, S., et al.\ 2005, The Astronomer's Telegram, 649,  

\bibitem[Foschini et al.(2004)]{foschini2004} Foschini, L., Tomsick, J.~A., Rodriguez, J., et al.\ 2004, 5th INTEGRAL Workshop on the INTEGRAL Universe, 552, 247 

\bibitem[Gim{\'e}nez-Garc{\'{\i}}a et al.(2015)]{gimenezgarcia2015} 
Gim{\'e}nez-Garc{\'{\i}}a, A., Torrej{\'o}n, J.~M., Eikmann, W., et al.\ 2015, Highlights of Spanish Astrophysics VIII, 482 

\bibitem[G\"uver \& Ozel(2009)]{guver2009} G\"uver, T. and Ozel, F.\ 2009, \mnras, 400, 2050

\bibitem[in 't Zand et al.(2003)]{intzand2003} in 't Zand, J.~J.~M., Ubertini, P., Capitanio, F., \& Del Santo, M.\ 2003, \iaucirc, 8077, 2 

\bibitem[Kallman et al.(2004)]{kallman2004} Kallman, T.~R., Palmeri, P., Bautista, M.~A., Mendoza, C., \& Krolik, J.~H.\ 2004, \apjs, 155, 675 

\bibitem[Krimm et al.(2013)]{krimm2013} Krimm, H.~A., Holland, S.~T., Corbet, R.~H.~D., et al.\ 2013, \apjs, 209, 14 

\bibitem[Lebrun et al.(2003)]{lebrun2003} Lebrun, F., Leray, J.~P., Lavocat, P., et al.\ 2003, \aap, 411, L141 

\bibitem[Lutovinov et al.(2005)]{lutovinov2005} Lutovinov, A., Rodriguez, J., Revnivtsev, M., \& Shtykovskiy, P.\ 2005, \aap, 433, L41 

\bibitem[Negueruela et al.(2006)]{negueruela2006} Negueruela, I., Smith, D.~M., Reig, P., Chaty, S., \& Torrej{\'o}n, J.~M.\ 2006, The X-ray Universe 2005, 604, 165 

\bibitem[Negueruela \& Schurch(2007)]{negueruela2007} Negueruela, I., \& Schurch, M.~P.~E.\ 2007, \aap, 461, 631 

\bibitem[Oskinova et al.(2012)]{oskinova2012} Oskinova, L.~M., Feldmeier, A., \& Kretschmar, P.\ 2012, \mnras, 421, 2820 

\bibitem[Rahoui et al.(2008)]{rahoui2008} Rahoui, F., Chaty, S., Lagage, P.-O., \& Pantin, E.\ 2008, \aap, 484, 801 

\bibitem[Rodriguez et al.(2003)]{rodriguez2003} Rodriguez, J., Tomsick, J.~A., Foschini, L., et al.\ 2003, \aap, 407, L41 

\bibitem[Rodriguez et al.(2006)]{rodriguez2006} Rodriguez, J., Bodaghee, A., Kaaret, P., et al.\ 2006, \mnras, 366, 274 

\bibitem[Sguera et al.(2006)]{sguera2006} Sguera, V., Bazzano, A., Bird, A.~J., et al.\ 2006, \apj, 646, 452 

\bibitem[Str{\"u}der et al.(2001)]{struder2001} Str{\"u}der, L., et al.\ 2001, \aap, 365, L18

\bibitem[Sugizaki et al.(2001)]{sugizaki2001} Sugizaki, M., Mitsuda, K., Kaneda, H., et al.\ 2001, \apjs, 134, 77 

\bibitem[Titarchuk(1994)]{titarchuk} Titarchuk, L.\ 1994, \apj, 434, 570 

\bibitem[Tomsick et al.(2003)]{tomsick2003} Tomsick, J.~A., Lingenfelter, R., Walter, R., et al.\ 2003, \iaucirc, 8076, 1 

\bibitem[Turner et al.(2001)]{turner2001} Turner, M.~J.~L., et al.\ 2001, \aap, 365, L27 

\bibitem[Ubertini et al.(2003)]{ubertini2003} Ubertini, P., Lebrun, F., Di Cocco, G., et al.\ 2003, \aap, 411, L131 

\bibitem[Walter et al.(2006)]{walter2006} Walter, R., Zurita Heras, J., Bassani, L., et al.\ 2006, A\&A, 453, 133

\bibitem[Wilms et al.(2000)]{wilms} Wilms, J., Allen, A., \& McCray, R.\ 2000, \apj, 542, 914 

\bibitem[Winkler et al.(2003)]{winkler2003} Winkler, C., Gehrels, N., Sch{\"o}nfelder, V., et al.\ 2003, \aap, 411, L349 

\bibitem[Zurita Heras et al.(2006)]{zuritaheras2006} Zurita Heras, J.~A., De Cesare, G., Walter, R., et al.\ 2006, \aap, 448, 261 

\bibitem[Zurita Heras et al.(2009)]{zuritaheras2009} Zurita Heras, J.~A., Chaty, S., Prat, L., \& Rodriguez, J.\ 2009, American Institute of Physics Conference Series, 1126, 313 

\end{thebibliography}
\end{document}